\documentclass[11pt,preprintnumbers,aps,amssymb,nofootinbib,amsmath,superscriptaddress,notitlepage,prd]
{revtex4-2}
\usepackage{epsfig,epsf}
\usepackage{comment}
\usepackage{bm} % puts greek and math symbols in boldface using \bm
\usepackage{color} % {\color{red} ... }
\usepackage{slashed} % Dirac slash
\usepackage{relsize}	%rescalable math 
\usepackage{soul} %spac­ing out (let­terspac­ing), un­der­lin­ing, strik­ing out, etc.
\usepackage{hyperref}
\usepackage{tensor} %\indices{^a_b} produces the properly spaced the tensor indices 
\usepackage{yfonts} %Gotic fonts \textgoth{This: is: Gothic}
\newcommand{\beq}{\begin{equation}}
\newcommand{\beql}[1]{\begin{equation}\label{#1}}
\newcommand{\eeq}{\end{equation}}
\def\bal#1\gal{\begin{align}#1\end{align}}
\newcommand{\ball}[1]{\bal\label{#1}}
\newcommand{\eq}[1]{(\ref{#1})}
\newcommand{\fig}[1]{Fig.~\ref{#1}}
\renewcommand{\sec}[1]{Sec.~\ref{#1}}
\DeclareMathOperator{\tr}{\mathrm{tr}}
\DeclareMathOperator{\Tr}{\mathrm{Tr}}

\renewcommand{\b}[1]{{\bm #1}} 
\usepackage{feynmp-auto}

\begin{document}

\title{Color Chiral Cherenkov radiation and energy loss in quark-gluon plasma}

\author{Jeremy Hansen}

\author{Kirill Tuchin}

\affiliation{
Department of Physics and Astronomy, Iowa State University, Ames, Iowa, 50011, USA}

\date{\today}

\begin{abstract}

We introduce and investigate the Color Chiral Cherenkov effect which consists in radiation of the circularly polarized gluons by a fast color charge moving with constant velocity in the presence of the Chiral Magnetic current. We derive the transition rates for all gluon polarizations. We compute the contribution of the Color Chiral Cherenkov effect to the parton energy loss in the quark-gluon plasma.

\end{abstract}

\maketitle

%%%%%%%%%%%%%%%%%%%%%%%%%%%%%%%%%%%%%%%%

\section{Introduction}\label{sec:intro}

A fast  particle carrying electric charge and moving in a medium with chiral fermions can lose a significant portion of its energy by means of the Chiral Cherenkov radiation \cite{Tuchin:2018sqe,Huang:2018hgk,Hansen:2020irw,Tuchin:2018mte}. It is emitted due to the unique dispersion relation of the electromagnetic field and is closely related to the chiral magnetic \cite{Kharzeev:2004ey,Kharzeev:2007jp,Kharzeev:2009fn,Kharzeev:2007tn,Fukushima:2008xe} and anomalous Hall effects \cite{Klinkhamer:2004hg,Zyuzin:2012tv,Grushin:2012mt,Kharzeev:2007tn,Kharzeev:2013ffa}. Thus far all studies of the Chiral Cherenkov radiation focused on the media governed by the Quantum Electrodynamics because of  possible technological  applications. However, it is clear that the color version of the Chiral Cherenkov effect is expected to give a significant contribution to the energy loss in strongly interacting media with chiral fermions. We therefore set in this paper to derive the color version of the Chiral Cherenkov effect and estimate its magnitude in the quark-gluon plasma. 

The color version of the chiral magnetic effect---the induction of the color current in the direction of the color magnetic field---can be described by letting the $\theta$-angle slowly vary with time. The gluon field excitations in the chiral medium supporting the chiral magnetic current are governed by the Lagrangian \cite{Wilczek:1987mv,Carroll:1989vb,Sikivie:1984yz}:
\ball{a1}
\mathcal{L}= -\frac{1}{2} \Tr\left(F_{\mu\nu} F^{\mu\nu}\right)-\frac{c_A}{2} \theta \Tr\left(  F_{\mu\nu}\tilde F^{\mu\nu}\right)  \,,
\gal
where the field tensor is
\ball{a1.1}
F_{\mu\nu}= \partial_\mu A_\nu -\partial_\nu A_\mu-ig[A_\mu, A_\nu]\,.
\gal
$A_\mu = A_\mu^at^a$ and $F_{\mu\nu}= F_{\mu\nu}^at^a$, where $t^a$ are the SU(3) generators. $\tilde F^{\mu\nu}= \frac{1}{2}\epsilon_{\mu\nu\lambda\rho} F^{\lambda\rho}$ is the dual field tensor. The external pseudo-scalar field $\theta$ is sourced by the topological charge and $c_A$ is the anomaly coefficient.

The equations of motion derived from the Lagrangian \eq{a1} depend only on the gradient $\partial_\mu \theta$. We adopt a model of the quark-gluon plasma with   spatially homogeneous and slowly time dependent $\theta$: $c_A\partial_\mu \theta= b_0\delta_{\mu 0}$, where $b_0$ is the (constant) chiral magnetic conductivity.
As a result the chiral magnetic current $\b j^a= b_0\b B^a$ emerges as a source of the color magnetic field in Amper's law. That $\theta$ is a slow function of time is indicated by the parametrically large 
sphaleron transition time $1/(g^4 T)$ at a given plasma temperature $T$ \cite{Arnold:1996dy,Arnold:1998cy,Bodeker:1998hm}.

The chiral magnetic current modifies the spectrum of the gluon excitations, which can be found by solving the equations of motion  without the self-interaction terms: 
\ball{a2}
\partial_\nu F^{\mu\nu}+b_\nu \tilde F^{\mu\nu}=0\,,
\gal
along with the Bianchi identity. In the radiation gauge, the corresponding vector potential obeys the equation
\ball{a3}
\partial_t^2\b A-\nabla^2\b A= b_0 (\b \nabla\times \b A)\,.
\gal
The plane wave solutions of \eq{a3} are circularly polarized and have the dispersion relation 
\ball{a4}
\omega^2= \b k^2 -\lambda b_0 |\b k|\,,
\gal
where $k=(\omega,\b k)$ and $\lambda=\pm 1$ indicate  the right or left polarization. With the account of the screening effects the
in-medium gluons have the dispersion relation 
\ball{t3}
\omega^2= \b k^2+\mu^2(\b k)= \b k^2 -\lambda b_0 |\b k|+\omega_p^2\,,
\gal
where $\mu$ is the gluon mass parameter and $\omega_p$ is the plasma frequency. Actually, Eq.~\eq{t3} is the short wavelength limit of the full dispersion relation \cite{Akamatsu:2013pjd,Manuel:2013zaa}, which nevertheless suffices for the present calculation.
Throughout the paper we assume that $b_0$ is positive. Since Eq.~\eq{t3} scales with $\lambda b_0$, the negative $b_0$ case can be obtained by the replacement $b_0\to -b_0$ and $\lambda\to -\lambda$.

The chiral term in the gluon dispersion relation generates the spacelike gluon mode  $\omega^2-\b k^2<0$ which opens the possibility for novel $1\to 2$ processes that are otherwise prohibited in QCD by energy and momentum conservation. This effect is analogous to the Cherenkov radiation where the spacelike excitations of the electromagnetic field produced by a particle moving at a speed greater than the phase velocity of light, represent the propagating wave solution in dielectric materials. In chiral media, the chiral conductivity effectively contributes to the medium dielectric response making possible excitation of the gauge field wave. This effect is referred to as the Chiral Cherenkov radiation. We refer to its QCD version as the Color Chiral Cherenkov radiation. 

\begin{figure}
    \centering  \includegraphics[width=.7\linewidth]{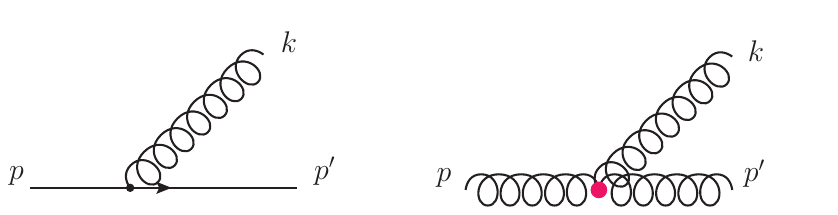}
    \caption{ $1\to 2$ processes contributing to the Color Chiral Cherenkov radiation. The anomalous contributions come about by the way of the gluon dispersion relation \eq{t3} and as an extra term in the  triple-gluon vertex \eq{f5}. The latter fact is indicated by the big red circle.
 }
    \label{fig:channels}
\end{figure}

Unlike QED, where the chrial Cherenkov radiation is described by a single diagram $e\to e\gamma$, where the photon dispersion relation is modified by the anomaly, in QCD there are two possible channels depicted in \fig{fig:channels}. The first of these channels $q\to qg$ is quasi-Abelian. The corresponding emission rate can be derived from the Abelian expression by including the appropriate color factors. We performed this calculation in 
\cite{Hansen:2020irw}. The novel channel is $g\to gg$ where all three gluons are excitations of the chiral medium. It is the focus of the present article. 

The paper is structured as follows. The next section deals with the Feynman rule for the triple-gluon vertex stemming from the second term in the right-hand-side of \eq{a1}. The main result, namely, the transition rates for all quark and gluon polarization states is derived in \sec{sec:rate}. Sec.~\ref{sec:loss} discusses the energy loss due to the Color Chiral Cherenkov radiation. The summary and outlook are presented in \sec{sec:sum}.

%%%%%%%%%%%%%%%%%%%%%%%%%%%%%%%%%%%%%%%%%%
\section{The triple-gluon vertex }\label{sec:vertex}

The diagrams contributing to the Color Chiral Cherenkov radiation are depicted in \fig{fig:channels}. Whereas the anomaly does not affect the Feynman rule of the fermion-gluon vertex, it does modify the triple-gluon one. To derive the rule we 
write in the second term in \eq{a1} 
\ball{f1}
\Tr\left( F^{\mu\nu}\tilde F^{\mu\nu}\right)= 2\partial_\mu K^\mu
\gal 
where
\ball{f2}
K^\mu= \epsilon^{\mu\nu\rho\sigma}\Tr\left(A_\nu\partial_\rho A_\sigma-g\frac{2i}{3}A_\nu A_\rho A_\sigma\right)
\gal
and integrate by parts the corresponding term in the action. The result is 
\ball{f4}
S_\theta= c_A\int  \partial_\mu\theta \epsilon^{\mu\nu\rho\sigma}\left(\frac{1}{2}A^a_\nu\partial_\rho A^a_\sigma-g\frac{2i}{3}\frac{1}{4}if^{abc}A^a_\nu A^b_\rho A^c_\sigma\right)d^4x\,.
\gal
The first term in \eq{f4} contributes to the equation of motion \eq{a2}, while the second one to the triple-gluon vertex. The variation of action $S_\theta$ with respect to three gluon fields produces the Feynman rule for the anomalous contribution to the triple-gluon vertex in momentum space:
\ball{f5}
\text{anomalous triple-gluon vertex}=gb_\mu\epsilon^{\mu\nu\rho\sigma}f^{abc}\,,
\gal
where we follow the conventions of  \cite{Peskin:1995ev}. This is the only new Feynman rule due to the anomalous term in the Lagrangian \eq{a1}. The red circle in \fig{fig:channels} includes both the conventional and anomalous contributions.

%Focusing on the three gluon term in \eq{f4} given three gluons with going into the vertex with momenta ${k_1}_\nu^a, {k_2}_\rho^b$ and ${k_3}_\sigma^c$ one finds that the anomalous vertex is equal to the following
%\ball{f6}
%\text{three gluon vertex}({k_1}_\nu^a, {k_2}_\rho^b,{k_3}_\sigma^c)=c_A  \partial_\mu\theta g f^{abc} \epsilon^{\mu\nu\rho\sigma}(2\pi)^4\delta^4(k_1+k_2+k_3)\,.
%\gal

%%%%%%%%%%%%%%%%%%%%%%%%%%%%%%%%%%%%%%%%%%
\section{The chiral Cherenkov radiation rate}\label{sec:rate}

The radiation rates can be computed as 
\ball{c9}
dW_{a\to bc}= \frac{g^2 C_{a\to bc}}{2(2\pi)^2}\sum_{ss'}\delta(\omega+E'-E)\delta(\b k+\b p'-\b p) \frac{1}{8E E' \omega}
  |\mathcal{M}_{a\to bc}|^2 d^3p'\, d^3k\,,
\gal
where the sum runs over the fermion polarization states, $\mathcal{M}_{a\to bc}$ are the amplitudes without color generators $t^a$ and the structure constants $f^{abc}$, and $C_{a\to bc}$ are the color factors. The color factors are given by
\bal
C_{q\to qg}&= \frac{1}{N_c}\tr(t^a t^a)=\frac{N_c^2-1}{2N_c}=\frac{4}{3}\,,\label{u1}\\
C_{g\to gg}&=\frac{(f^{abc})^2}{N_c^2-1}=N_c=3\,.\label{u2}
\gal
The matrix elements read:
\bal
i\mathcal{M}_{q\to qg}&=i\bar u_{\b p' s'}\slashed{e}^*_{\b k \lambda} u_{\b p s}\,,\label{t1}\\
i\mathcal{M}_{g\to gg}&=i\mathcal{M}^A_{g\to gg}+ i\mathcal{M}^B_{g\to gg}\nonumber\\
&=
(e_{\b p \lambda_0}\cdot e^*_{\b k \lambda})(p+k)\cdot e^*_{\b p' \lambda'}
+(e^*_{\b k \lambda}\cdot e^*_{\b p' \lambda'})(p'-k)\cdot e_{\b p \lambda_0}-
(e_{\b p \lambda_0}\cdot e^*_{\b p' \lambda'})(p+p')\cdot e^*_{\b k \lambda}\nonumber\\
&-ib_\mu\epsilon^{\mu\nu\rho\sigma}e_{\b p \lambda_0,\nu}e^*_{\b k \lambda,\rho}e^*_{\b p' \lambda',\sigma}\,,\label{t2}
\gal
where $e_{\b p\lambda \mu}$'s are the circular polarization vectors and the superscripts $A$,$B$ refer to the second and third lines of \eq{t2} respectively. Note that not just the last line in \eq{t2}, but all terms in \eq{t1} and \eq{t2} depend on the anomaly by the way of the dispersion relation \eq{t3}.

We perform the calculation in the high-energy approximation, meaning that the momenta of all particles along the jet  axis are much larger than the transverse momenta and the effective mass. Let $z$ be the jet axis, then 
\begin{subequations}\label{C1a}
\bal
&p=\left(E,0,0,E-\frac{k_\perp^2+x(1-x)\mu^2(E)}{2x(1-x)E}\right)\,,\label{C1aa}\\
&k=\left(\omega,k_\perp,0,\omega-\frac{k_\perp^2+\mu^2(\omega)}{2\omega}\right)\,,\label{C1ab}\\
&p'=\left(E',-k_\perp,0,E'-\frac{k_\perp^2+\mu^2(E')}{2E'}\right) \,,\label{C1ac}
\gal
\end{subequations}
where the mass parameter in \eq{t3} becomes $\mu^2(\omega)\approx \omega_p^2-\lambda b_0 \omega$ and $x=\omega/E$ is the fraction of energy of the incident parton carried away by the gluon $k$. Eqs.~\eq{C1a} assume that $k_\bot \ll E,E',\omega$ and $|\mu|\ll  E,E',\omega$. The latter condition is equivalent to $b_0\ll E,E',\omega$.

Given \eq{C1a}, the polarization vectors can be chosen as 
\begin{subequations}\label{C1b}
\bal
&e_{p \lambda_0}=\frac{1}{\sqrt{2}}(0,1,\lambda_0 i,0)\,,\label{C1ba}\\
&e_{k \lambda}=\frac{1}{\sqrt{2}}\left(0,1,\lambda i,-\frac{k_\perp}{\omega}\right)\,,\label{C1bb}\\
&e_{p' \lambda'}=\frac{1}{\sqrt{2}}\left(0,1,\lambda' i,\frac{k_\perp}{E'}\right)\,,\label{C1bc}
\gal
\end{subequations}

The integral over the $\b p'$ in \eq{c9} is trivial considering the delta-function expressing the momentum conservation. The remaining delta-function in $q\to qg$ rate reads:
\ball{t10}
\delta(E'+\omega-E)&=2x(1-x)E\delta\left[k_\bot^2+\mu^2(\omega)(1-x)+m^2x^2\right]\,,\nonumber\\
&=2x(1-x)E\delta\left[k_\bot^2+(\omega_p^2-\lambda b_0 xE)(1-x)+m^2x^2\right]\,.
\gal
Clearly, the argument of this delta-function can vanish only if $\lambda b_0>0$. We assumed in Introduction that $b_0>0$. Therefore, only the right-polarized gluons $\lambda>0$ can be radiated by the incident quark. Moreover, the energy conservation, expressed by the argument of the delta-function \eq{t10}, can be satisfied only if 
\ball{t12}
x_-^{q\to qg} <x<x_+^{q\to qg}\,,
\gal
where 
\ball{t13}
x_\pm^{q\to qg}=\frac{\omega_p^2+\lambda b_0 E\pm\sqrt{(\omega_p^2-\lambda b_0 E)^2-4m^2\omega_p^2}}{2(m^2+\lambda b_0 E)}\,.
\gal
The requirement  that \eq{t13} have real values sets the infrared threshold for the energy of the incident quark:
\ball{t14}
E>E_1=\frac{2m\omega_p+\omega_p^2}{b_0}\,.
\gal

The  delta-function in $g\to gg$ rate reads:
\ball{t15}
\delta(E'+\omega-E)&=2x(1-x)E\delta\left[k_\perp^2+x\mu^2(E')+(1-x)\mu^2(\omega)-x(1-x)\mu^2(E)\right]\nonumber\\
&=2x(1-x)E\delta\left[k_\perp^2+\omega_p^2(1-x+x^2)-b_0E(\lambda+\lambda'-\lambda_0)x(1-x)\right]\,.
\gal
In this case the gluon emission is possible only if $\lambda+\lambda'>\lambda_0$. Apparently only the following four channels are allowed: $g_L\to g_Rg_L$, $g_L\to g_Lg_R$, $g_R\to g_Rg_R$, $g_L\to g_Rg_R$. In the first three 
 of these channels $\lambda+\lambda'-\lambda_0=1$, whereas in the last one $\lambda+\lambda'-\lambda_0=3$. Additionally,  
\ball{t18}
x_-^{g\to gg} <x<x_+^{g\to gg}\,,
\gal
where 
\ball{t19}
x_\pm^{g\to gg}=\frac{\omega_p^2+(\lambda+\lambda'-\lambda_0) b_0 E\pm\sqrt{(\omega_p^2-(\lambda+\lambda'-\lambda_0) b_0 E)^2-4\omega_p^4}}{2(\omega_p^2+(\lambda+\lambda'-\lambda_0) b_0 E)}\,.
\gal
The requirement  that \eq{t19} have real values sets the infrared threshold for the energy of the incident gluon:
\ball{t21}
E>E_2=\frac{3\omega_p^2}{b_0}\,.
\gal

%Eqs.~\eq{t13},\eq{t19} simplify in the limit $b_0\gg \omega_p$ 
%\begin{subequations}\label{C12}
%\bal
%&x^{q\to qg}_\pm=\frac{\lambda b_0 E\pm\left(\lambda b_0 E-2\frac{\omega_p^2}{\lambda b_0 E}m^2\right)}{2(\lambda b_0 E+m^2)}\,,\label{C12a}\\
%&x^{g\to gg}_\pm=\frac{1\pm(1-\frac{\omega_p^2}{(\lambda+\lambda'-\lambda_0)b_0E})}{2} \,.\label{C12c}
%\gal
%\end{subequations}

The amplitude $\mathcal{M}_{q\to qg}$ was computed in \cite{Tuchin:2018sqe}:
\bal
\sum_{ss'}|\mathcal{M}_{q\to qg_R}|^2&=4\left[E E'-m^2-\frac{( \b k\cdot{\b  p})(\b k\cdot{\b p'})}{\b k^2}\right]\,,\label{t5}\\
\sum_{ss'}|\mathcal{M}_{q\to qg_L}|^2&=0\,.\label{t5b}
\gal
For the gluon splitting amplitudes we find
\begin{subequations}\label{C1c}
\bal
&i\mathcal{M}^A_{g_R\rightarrow g_R g_R}=\frac{k_\perp}{x(1-x)}\,,\label{C1ca}\\
&iM^A_{g_L\rightarrow g_R g_L}=\frac{(1-x) k_\perp}{x}\,,\label{C1cb}\\
&iM^A_{g_L\rightarrow g_L g_R}=\frac{x k_\perp}{(1-x)}\,,\label{C1cc}\\
&iM^A_{g_L\rightarrow g_R g_R}=\mathcal{O}(k_\perp^2/E, b_0/E)\,,\label{C1bd}\\
&i M^B_{g\rightarrow g g}=\frac{-b_0 k_\perp(\lambda(1-x)+\lambda'x+\lambda_0)}{x(1-x)E}\,.\label{C1d}
\gal
\end{subequations}
provided that $k_\bot,b_0\ll E,E',\omega$. In the high energy limit $\mathcal{M}^B_{g\rightarrow g g}$ appears only as a sub-leading correction to $\mathcal{M}^A_{g\rightarrow g g}$ and is therefore can be neglected in our calculation. We note incidentally, that $\mathcal{M}^B_{g_L\rightarrow g_R g_R}=0$ so that the anomalous vertex correction does not contribute even in the sub-leading channel \eq{C1bd}. 
Substituting \eq{t5},\eq{C1c},\eq{u1},\eq{u2} into \eq{c9} and summing over the final gluon polarizations we derive 
\begin{subequations}\label{C2}
\bal
&\frac{dW_{q\rightarrow qg}}{d k_\perp^2 d x}=\frac{\alpha_s g^2}{3x^2 (1-x)E }\left[(1+(1-x)^2) k_\perp^2+m^2x^4\right]\delta\left(k_\perp^2+\mu^2 (1-x)+m^2x^2\right) \,,\label{C2a}\\
&\frac{dW_{g_R\rightarrow g g}}{d k_\perp^2 d x}=\frac{3\alpha_s g^2}{2x^2 (1-x)^2E } k_\perp^2\delta\left(k_\perp^2+\omega_p^2x^2+(\omega_p^2-b_0 \omega) (1-x)\right) \,,\label{C2aa}\\
&\frac{dW_{g_L\rightarrow g g}}{d k_\perp^2 d x}=\frac{3\alpha_s g^2 }{2 E}\left[\frac{ (1-x)^2}{x^2 }+\frac{ x^2}{(1-x)^2 }\right] k_\perp^2\delta\left(k_\perp^2+\omega_p^2x^2+(\omega_p^2-b_0 \omega) (1-x)\right) \,.\label{C2ab}
\gal
\end{subequations}
One can easily identify in \eq{C2} the contributions to the standard splitting functions $P_{gq}(x)$ and $P_{gg}(x)$ corresponding to gluon emission.
Integrating over the transverse momentum  $k_\bot$ we obtain the spectra of the gluon emission rate: 
\begin{subequations}\label{C11}
\bal
\frac{dW_{q\rightarrow qg}}{d x}=\frac{\alpha_s g^2}{3 x^2 E}\left\{\left[1+(1-x)^2\right] \left(b_0 xE-\omega_p^2\right)- 2 m^2 x^2\right\}\Theta\left(x^{q\to qg}_+-x\right)\Theta\left(x-x^{q\to qg}_-\right)  \,,\label{C11a}\\
%&\frac{dW_{g_R\rightarrow gg}}{d x}=\frac{3\alpha_s g^2}{4 x^2(1-x)^2E }\left\{\left(b_0 E x-\omega_p^2\right) (1-x)-\omega_p^2x^2\right\}\,\Theta\left(x^{g\to gg}_+-x\right)\Theta\left(x-x^{g\to gg}_-\right) \,,\label{C11b}\\
%&\frac{dW_{g_L\rightarrow gg}}{d x}=\frac{3\alpha_s g^2}{4x^2 (1-x)^2E }\left\{\left(b_0 E x-\omega_p^2\right) (1-x)-\omega_p^2x^2\right\}\left[x^4+(1-x)^4\right] \Theta\left(x^{g\to gg}_+-x\right)\Theta\left(x-x^{g\to gg}_-\right)  \,.\label{C11c}\\
\frac{dW_{g_{\lambda_0}\rightarrow gg}}{d x}=\frac{3\alpha_s g^2}{4x^2 (1-x)^2E }\left\{\left(b_0 E x-\omega_p^2\right) (1-x)-\omega_p^2x^2\right\} \left\{ \left[x^4+(1-x)^4\right]\delta_{\lambda_0,-1}+\delta_{\lambda_0,1} \right\} \nonumber \\
\times\Theta\left(x^{g\to gg}_+-x\right)\Theta\left(x-x^{g\to gg}_-\right)
\,.\label{C11d}
\gal
\end{subequations}
Here $\Theta$ is the Heaviside step-function, and $\lambda_0=\pm 1$ is the right/left-hand polarizations of the incident gluon.

%%%%%%%%%%%%%%%%%%%%%%%%%%%%%%%%%%%%%%
%%%%%%%%%%%%%%%%%%%%%%%%%%%%%%%%%%%%%%%
\section{Color chiral Cherenkov radiation in quark-gluon plasma}\label{sec:loss}

\begin{figure}[ht]
\begin{tabular}{cc}
      \includegraphics[width=0.5\linewidth]{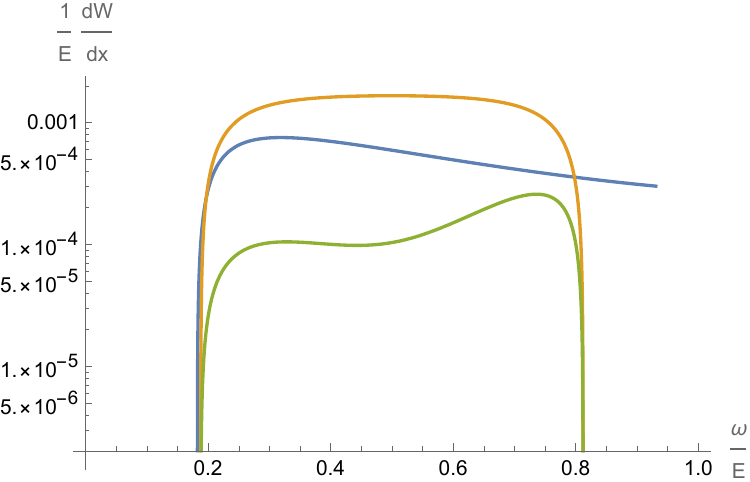} &
      \includegraphics[width=0.5\linewidth]{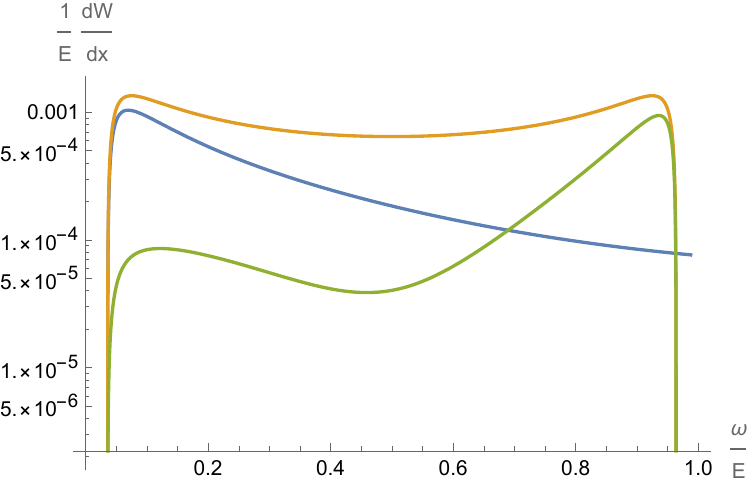}  
      \end{tabular}
  \caption{The Color Chiral Cherenkov radiation rates versus $x=\omega/E$. 
  The channels $q\rightarrow q g$, $g_R\rightarrow g g$, and $g_L\rightarrow g g$ are represented by blue, orange, and green lines respectively. Left panel: $E=20$~GeV,  right panel: $E=100$~GeV. Both panels: $g=2$, $T=300$~MeV, $b_0=50$~MeV.}
\label{fig:rateplot}
\end{figure}

We now consider the rate of gluon radiation due to the color chiral magnetic current in quark-gluon plasma. The plasma frequency and the quark thermal mass are given by $\omega_p^2=\frac{g^2T^2}{18}(2N_c+N_f)$ and  $m^2=\frac{g^2T^2}{16N_c}(N_c^2-1)$ respectively \cite{RISCHKE2004197}. Using these equations in \eq{C11} we compute the anomalous contribution to the gluon emission spectra from quark-gluon plasma shown in \fig{fig:rateplot}. We emphasize that this calculation takes into account only the anomaly--induced radiation which is but a fraction of the total gluon radiation.  
%%%%%
One can observe that the gluon spectrum is constrained by the thresholds \eq{t13},\eq{t19}. The thresholds nearly coincide in the infrared, but are different in the ultraviolet part of the spectrum. The right-handed gluons decay most readily except in the far infrared which is dominated by the radiation off the quark.

%%%%%%%%%%%%%%%%%%%%%%%%%%%%%%%%%%%%%%
%%%%%%%%%%%%%%%%%%%%%%%%%%%%%%%%%%%%%%%
%\section{Energy loss}\label{sec:loss}

\begin{figure}[ht]
\begin{tabular}{cc}
      \includegraphics[width=0.7\linewidth]{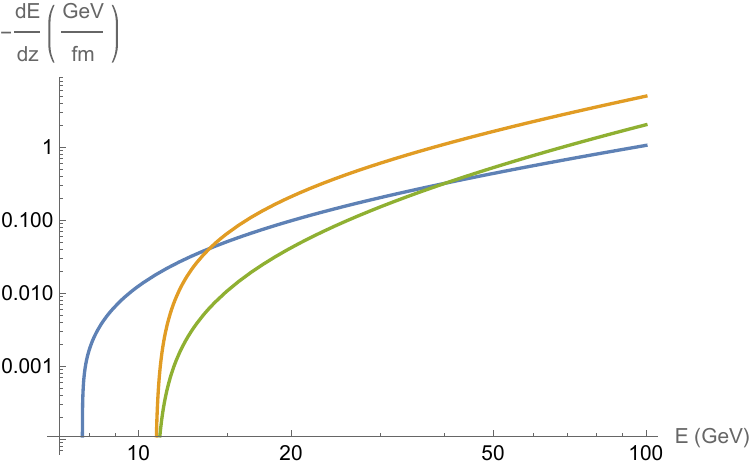}   
      \end{tabular}
  \caption{The rate of energy loss due to the Color Chiral Cherenkov radiation for $q\rightarrow q g$ (blue), $g_R\rightarrow g g$ (orange), and $g_L\rightarrow g g$ (green). Parameters: $g=2$, $T=300$~MeV, $b_0=50$~MeV.  }
\label{fig:loss}
\end{figure}

The contribution of the Color Chiral Cherenkov radiation to the energy loss by a fast parton propagating in quark-gluon plasma along the $z$-axis is given by 
\ball{t20}
-\frac{dE_{a\to bc}}{dz}=\int_0^E \omega \frac{dW_{a\to bc}}{d\omega} d\omega = E\int_0^1 x\frac{dW_{a\to bc}}{dx}dx\,.
\gal
Substituting \eq{C11} into \eq{t20} yields for each channel 
\begin{subequations}\label{C9a}
\bal
-\frac{dE_{q\rightarrow qg}}{d z}=&\frac{2\alpha_s g^2}{ 3}\bigg\{(x_+^{q\rightarrow q g}-x_-^{q\rightarrow q g})\left[(x_+^{q\rightarrow q g}+x_-^{q\rightarrow q g}-8)b_0 E-8m^2-2\omega_p^2\right]\nonumber\\
   & +(b_0 E+\omega_p^2)\ln{\frac{x_+^{q\rightarrow q g}}{x_-^{q\rightarrow q g}}}\bigg\}\Theta\left(E-E_1\right)  \,,\label{C9aa1}\\
-\frac{dE_{g_R\rightarrow gg}}{d z}=&\frac{3\alpha_s g^2}{4 }\bigg\{\left(b_0 E-\omega_p^2\right)\ln{\frac{x_+^{g\rightarrow gg}}{x_-^{g\rightarrow gg}}}
-2\left(b_0 E+\omega_p^2 \right)(x_+^{g\rightarrow gg}-x_-^{g\rightarrow gg})\bigg\}\Theta\left(E-E_2\right) \,,\label{C9ab}\\
-\frac{dE_{g_L\rightarrow gg}}{d z} =&\frac{\alpha_s g^2}{4}\bigg\{\frac{3\left[(b_0 E+2\omega_p^2)^2-\omega_p^4\right]}{b_0 E+\omega_p^2}\ln{\frac{x_+^{g\rightarrow gg}}{x_-^{g\rightarrow gg}}}\nonumber\\
&-(17b_0 E+21\omega_p^2)(x_+^{g\rightarrow gg}-x_-^{g\rightarrow gg})\bigg\}\Theta\left(E-E_2\right)\,.\label{C9ac}
\gal
\end{subequations}In the high energy limit $b_0 E\gg m^2,\omega_p^2$ equations \eq{C9a} simplify as follows:
\begin{subequations}\label{C9b}
\bal
&-\frac{dE_{q\rightarrow qg}}{d z}=\frac{4\alpha_s g^2b_0E}{ 9}\,,\label{C9ba}\\
&-\frac{dE_{g_R\rightarrow gg}}{d z}=\frac{3\alpha_s g^2b_0 E}{4 }\left(\ln{\frac{b_0 E}{\omega_p^2}}-1\right) \,,\label{C9bb}\\
&-\frac{dE_{g_L\rightarrow gg}}{d z} =\frac{3\alpha_s g^2b_0 E}{4}\left(\ln{\frac{b_0 E}{\omega_p^2}}-\frac{17}{6}\right)\,.\label{C9bc}
\gal
\end{subequations}
\fig{fig:loss} exhibits the contribution of the Color Chiral Cherenkov radiation to the parton energy loss.  We observe in that at high energy the right-hand gluon loses more energy than the left-handed one and the quark, while at lower energy the quark channel is the main mechanism of energy loss. 

The energy loss of a jet consisting of many $q$,$\bar q$,$g$ states is determined by solving the system of coupled evolution equations for the parton distribution functions. Suppose that $f_a(x)$ are the momentum distribution functions of a species $a=q,\bar q,g$ normalized as
\ball{s1}
\sum_a\int_0^1 f_a(x)xdx=1\,.
\gal
Then, the jet energy loss is
\ball{s2}
-\frac{dE}{dz}=\sum_a\int_0^1 dx f_a(x) \left(-\frac{dE_a}{dz}\right)\,,
\gal
where the partial energy losses are given by Eqs.~\eq{C9a}. Of course, jet energy loss due to the chiral anomaly is only a fraction of the total energy loss and therefore its phenomenological significance can be assessed only by incorporating the anomaly--induced corrections into the standard evolution equations.

%%%%%%%%%%%%%%%%%%%%%%%%%%%%%%%%%%%%%%
%%%%%%%%%%%%%%%%%%%%%%%%%%%%%%%%%%%%%%%
\section{Summary and outlook}\label{sec:sum}

We  investigated the Color Chiral Cherenkov radiation---the QCD  analogue of the Chiral Cherenkov effect in QED---and applied the results to study its role in the energy loss and jet quenching in the quark-gluon plasma. Our main result is 
the gluon emission rates \eq{C11} by quark and gluons in the chiral medium. The resulting energy loss is given by Eqs.~\eq{C9a}. 

A number of approximations were made to derive these results. Firstly, we assumed that the medium is homogeneous such that $\b b=c_A\b \nabla \theta=0$. One can consider a complimentary scenario where  $\b b$ is finite and $b_0$ vanishes. The corresponding rate of gluon emission by a fast quark can be obtained similarly to QED, by substituting $b_0\rightarrow |\b b| \cos{\beta}$ where $\beta$ into \eq{C11} is the angle between $\b b$ and the outgoing gluon \cite{Tuchin:2018sqe,Tuchin:2018mte}. Gluon radiation by a fast gluon at finite $\b b$, on the other hand, requires further analysis. However, it should be noted that while $\mathcal{M}^B_{g\to gg}$ gives a sub-relativistic correction in a homogeneous medium, it gives no contribution at all in the inhomogeneous case. Indeed, the scattering amplitude $\mathcal{M}^B_{g\to gg}=-ib_\mu\epsilon^{\mu\nu\rho\sigma}e_{\b p \lambda_0,\nu}e^*_{\b k \lambda,\rho}e^*_{\b p' \lambda',\sigma}$ is antisymmetric with respect to exchange of $b_\mu$ and any polarization vector. As such if $b_0=0$, then $\mathcal{M}^B_{g\to gg}$ must be proportional to the zeroth component of one of the polarization vectors. Therefore, given that the zeroth component is zero (see \eq{C1b}), $\mathcal{M}^B_{g\to gg}$  vanishes.

Secondly, we assumed that the radiating particle is ultra-relativistic throughout the entire process. As a result, the spectrum in the quark channel exhibits discontinuous behavior at  $x=x^{q\to qg}_+$, as can be seen in \fig{fig:rateplot}.  By including non-relativistic corrections for the outgoing particle, one may expect the rate of gluon emission to drop smoothly to zero as $x$ tends to 1. This correction has very little effect on the rate of energy loss, however it may play a role in parton evolution of jet.

 Comparing the decay rates of the left-handed and right-handed gluons in \fig{fig:rateplot}, one finds that in a medium with $b_0>0$, the left-handed gluons decay more slowly when compared to right-handed ones. As a result jets develop strong left-hand polarziation. This is the clearest manifestation of the chiral anomaly in the jet structure. 
 
 The chiral imbalance is also imprinted into the jet loss pattern seen in  \fig{fig:loss}: the right-handed gluons loose a lot more energy than the left-handed ones.  However, as the jet energy increases, the difference between the gluon polarizations becomes less pronounced since the energy loss is driven primarily by the large polarization independent logarithms in \eq{C9b}. These logarithms also enhance the relative contribution of gluons as compared to quarks to the energy loss at high energies. In contrast, at low energy, the energy loss is dominated by quarks as seen in \fig{fig:loss}. We stress again that these conclusions hold only for energy loss due to the Color Chiral Cherenkov radiation, ignored all other contributions. A recent review of the conventional mechanisms of energy loss can be found in \cite{Cao_2021}. 
 
In conclusion, energy loss due to the Color Chiral Cherenkov radiation is significant. Therefore a comprehensive phenomenological analysis requires incorporation of the novel energy loss channels into the numerical packages describing jets in hot nuclear medium.

%%%%%%%%%%%%%%%%%%%%%%%%%%%%%%%%
\acknowledgments
This work  was supported in part by the U.S. Department of Energy Grants No.\ DE-SC0023692.

%%%%%%%%%%%%
\bibliographystyle{apsrev4-2}
\bibliography{anom-biblio}

\begin{thebibliography}{24}%
\makeatletter
\providecommand \@ifxundefined [1]{%
 \@ifx{#1\undefined}
}%
\providecommand \@ifnum [1]{%
 \ifnum #1\expandafter \@firstoftwo
 \else \expandafter \@secondoftwo
 \fi
}%
\providecommand \@ifx [1]{%
 \ifx #1\expandafter \@firstoftwo
 \else \expandafter \@secondoftwo
 \fi
}%
\providecommand \natexlab [1]{#1}%
\providecommand \enquote  [1]{``#1''}%
\providecommand \bibnamefont  [1]{#1}%
\providecommand \bibfnamefont [1]{#1}%
\providecommand \citenamefont [1]{#1}%
\providecommand \href@noop [0]{\@secondoftwo}%
\providecommand \href [0]{\begingroup \@sanitize@url \@href}%
\providecommand \@href[1]{\@@startlink{#1}\@@href}%
\providecommand \@@href[1]{\endgroup#1\@@endlink}%
\providecommand \@sanitize@url [0]{\catcode `\\12\catcode `\$12\catcode
  `\&12\catcode `\#12\catcode `\^12\catcode `\_12\catcode `\%12\relax}%
\providecommand \@@startlink[1]{}%
\providecommand \@@endlink[0]{}%
\providecommand \url  [0]{\begingroup\@sanitize@url \@url }%
\providecommand \@url [1]{\endgroup\@href {#1}{\urlprefix }}%
\providecommand \urlprefix  [0]{URL }%
\providecommand \Eprint [0]{\href }%
\providecommand \doibase [0]{https://doi.org/}%
\providecommand \selectlanguage [0]{\@gobble}%
\providecommand \bibinfo  [0]{\@secondoftwo}%
\providecommand \bibfield  [0]{\@secondoftwo}%
\providecommand \translation [1]{[#1]}%
\providecommand \BibitemOpen [0]{}%
\providecommand \bibitemStop [0]{}%
\providecommand \bibitemNoStop [0]{.\EOS\space}%
\providecommand \EOS [0]{\spacefactor3000\relax}%
\providecommand \BibitemShut  [1]{\csname bibitem#1\endcsname}%
\let\auto@bib@innerbib\@empty
%</preamble>
\bibitem [{\citenamefont {Tuchin}(2018{\natexlab{a}})}]{Tuchin:2018sqe}%
  \BibitemOpen
  \bibfield  {author} {\bibinfo {author} {\bibfnamefont {K.}~\bibnamefont
  {Tuchin}},\ }\href {https://doi.org/10.1016/j.physletb.2018.09.055}
  {\bibfield  {journal} {\bibinfo  {journal} {Phys. Lett. B}\ }\textbf
  {\bibinfo {volume} {786}},\ \bibinfo {pages} {249} (\bibinfo {year}
  {2018}{\natexlab{a}})},\ \Eprint {https://arxiv.org/abs/1806.07340}
  {arXiv:1806.07340 [hep-ph]} \BibitemShut {NoStop}%
\bibitem [{\citenamefont {Huang}\ and\ \citenamefont
  {Tuchin}(2018)}]{Huang:2018hgk}%
  \BibitemOpen
  \bibfield  {author} {\bibinfo {author} {\bibfnamefont {X.-G.}\ \bibnamefont
  {Huang}}\ and\ \bibinfo {author} {\bibfnamefont {K.}~\bibnamefont {Tuchin}},\
  }\href {https://doi.org/10.1103/PhysRevLett.121.182301} {\bibfield  {journal}
  {\bibinfo  {journal} {Phys. Rev. Lett.}\ }\textbf {\bibinfo {volume} {121}},\
  \bibinfo {pages} {182301} (\bibinfo {year} {2018})},\ \Eprint
  {https://arxiv.org/abs/1808.00635} {arXiv:1808.00635 [hep-ph]} \BibitemShut
  {NoStop}%
\bibitem [{\citenamefont {Hansen}\ and\ \citenamefont
  {Tuchin}(2021)}]{Hansen:2020irw}%
  \BibitemOpen
  \bibfield  {author} {\bibinfo {author} {\bibfnamefont {J.}~\bibnamefont
  {Hansen}}\ and\ \bibinfo {author} {\bibfnamefont {K.}~\bibnamefont
  {Tuchin}},\ }\href {https://doi.org/10.1103/PhysRevC.104.034903} {\bibfield
  {journal} {\bibinfo  {journal} {Phys. Rev. C}\ }\textbf {\bibinfo {volume}
  {104}},\ \bibinfo {pages} {034903} (\bibinfo {year} {2021})},\ \Eprint
  {https://arxiv.org/abs/2012.06089} {arXiv:2012.06089 [hep-ph]} \BibitemShut
  {NoStop}%
\bibitem [{\citenamefont {Tuchin}(2018{\natexlab{b}})}]{Tuchin:2018mte}%
  \BibitemOpen
  \bibfield  {author} {\bibinfo {author} {\bibfnamefont {K.}~\bibnamefont
  {Tuchin}},\ }\href {https://doi.org/10.1103/PhysRevD.98.114026} {\bibfield
  {journal} {\bibinfo  {journal} {Phys. Rev. D}\ }\textbf {\bibinfo {volume}
  {98}},\ \bibinfo {pages} {114026} (\bibinfo {year} {2018}{\natexlab{b}})},\
  \Eprint {https://arxiv.org/abs/1809.08181} {arXiv:1809.08181 [hep-ph]}
  \BibitemShut {NoStop}%
\bibitem [{\citenamefont {Kharzeev}(2006)}]{Kharzeev:2004ey}%
  \BibitemOpen
  \bibfield  {author} {\bibinfo {author} {\bibfnamefont {D.}~\bibnamefont
  {Kharzeev}},\ }\href {https://doi.org/10.1016/j.physletb.2005.11.075}
  {\bibfield  {journal} {\bibinfo  {journal} {Phys. Lett. B}\ }\textbf
  {\bibinfo {volume} {633}},\ \bibinfo {pages} {260} (\bibinfo {year}
  {2006})},\ \Eprint {https://arxiv.org/abs/hep-ph/0406125}
  {arXiv:hep-ph/0406125} \BibitemShut {NoStop}%
\bibitem [{\citenamefont {Kharzeev}\ \emph {et~al.}(2008)\citenamefont
  {Kharzeev}, \citenamefont {McLerran},\ and\ \citenamefont
  {Warringa}}]{Kharzeev:2007jp}%
  \BibitemOpen
  \bibfield  {author} {\bibinfo {author} {\bibfnamefont {D.~E.}\ \bibnamefont
  {Kharzeev}}, \bibinfo {author} {\bibfnamefont {L.~D.}\ \bibnamefont
  {McLerran}},\ and\ \bibinfo {author} {\bibfnamefont {H.~J.}\ \bibnamefont
  {Warringa}},\ }\href {https://doi.org/10.1016/j.nuclphysa.2008.02.298}
  {\bibfield  {journal} {\bibinfo  {journal} {Nucl. Phys. A}\ }\textbf
  {\bibinfo {volume} {803}},\ \bibinfo {pages} {227} (\bibinfo {year}
  {2008})},\ \Eprint {https://arxiv.org/abs/0711.0950} {arXiv:0711.0950
  [hep-ph]} \BibitemShut {NoStop}%
\bibitem [{\citenamefont {Kharzeev}(2010)}]{Kharzeev:2009fn}%
  \BibitemOpen
  \bibfield  {author} {\bibinfo {author} {\bibfnamefont {D.~E.}\ \bibnamefont
  {Kharzeev}},\ }\href {https://doi.org/10.1016/j.aop.2009.11.002} {\bibfield
  {journal} {\bibinfo  {journal} {Annals Phys.}\ }\textbf {\bibinfo {volume}
  {325}},\ \bibinfo {pages} {205} (\bibinfo {year} {2010})},\ \Eprint
  {https://arxiv.org/abs/0911.3715} {arXiv:0911.3715 [hep-ph]} \BibitemShut
  {NoStop}%
\bibitem [{\citenamefont {Kharzeev}\ and\ \citenamefont
  {Zhitnitsky}(2007)}]{Kharzeev:2007tn}%
  \BibitemOpen
  \bibfield  {author} {\bibinfo {author} {\bibfnamefont {D.}~\bibnamefont
  {Kharzeev}}\ and\ \bibinfo {author} {\bibfnamefont {A.}~\bibnamefont
  {Zhitnitsky}},\ }\href {https://doi.org/10.1016/j.nuclphysa.2007.10.001}
  {\bibfield  {journal} {\bibinfo  {journal} {Nucl. Phys. A}\ }\textbf
  {\bibinfo {volume} {797}},\ \bibinfo {pages} {67} (\bibinfo {year} {2007})},\
  \Eprint {https://arxiv.org/abs/0706.1026} {arXiv:0706.1026 [hep-ph]}
  \BibitemShut {NoStop}%
\bibitem [{\citenamefont {Fukushima}\ \emph {et~al.}(2008)\citenamefont
  {Fukushima}, \citenamefont {Kharzeev},\ and\ \citenamefont
  {Warringa}}]{Fukushima:2008xe}%
  \BibitemOpen
  \bibfield  {author} {\bibinfo {author} {\bibfnamefont {K.}~\bibnamefont
  {Fukushima}}, \bibinfo {author} {\bibfnamefont {D.~E.}\ \bibnamefont
  {Kharzeev}},\ and\ \bibinfo {author} {\bibfnamefont {H.~J.}\ \bibnamefont
  {Warringa}},\ }\href {https://doi.org/10.1103/PhysRevD.78.074033} {\bibfield
  {journal} {\bibinfo  {journal} {Phys. Rev. D}\ }\textbf {\bibinfo {volume}
  {78}},\ \bibinfo {pages} {074033} (\bibinfo {year} {2008})},\ \Eprint
  {https://arxiv.org/abs/0808.3382} {arXiv:0808.3382 [hep-ph]} \BibitemShut
  {NoStop}%
\bibitem [{\citenamefont {Klinkhamer}\ and\ \citenamefont
  {Volovik}(2005)}]{Klinkhamer:2004hg}%
  \BibitemOpen
  \bibfield  {author} {\bibinfo {author} {\bibfnamefont {F.~R.}\ \bibnamefont
  {Klinkhamer}}\ and\ \bibinfo {author} {\bibfnamefont {G.~E.}\ \bibnamefont
  {Volovik}},\ }\href {https://doi.org/10.1142/S0217751X05020902} {\bibfield
  {journal} {\bibinfo  {journal} {Int. J. Mod. Phys. A}\ }\textbf {\bibinfo
  {volume} {20}},\ \bibinfo {pages} {2795} (\bibinfo {year} {2005})},\ \Eprint
  {https://arxiv.org/abs/hep-th/0403037} {arXiv:hep-th/0403037} \BibitemShut
  {NoStop}%
\bibitem [{\citenamefont {Zyuzin}\ and\ \citenamefont
  {Burkov}(2012)}]{Zyuzin:2012tv}%
  \BibitemOpen
  \bibfield  {author} {\bibinfo {author} {\bibfnamefont {A.~A.}\ \bibnamefont
  {Zyuzin}}\ and\ \bibinfo {author} {\bibfnamefont {A.~A.}\ \bibnamefont
  {Burkov}},\ }\href {https://doi.org/10.1103/PhysRevB.86.115133} {\bibfield
  {journal} {\bibinfo  {journal} {Phys. Rev. B}\ }\textbf {\bibinfo {volume}
  {86}},\ \bibinfo {pages} {115133} (\bibinfo {year} {2012})},\ \Eprint
  {https://arxiv.org/abs/1206.1868} {arXiv:1206.1868 [cond-mat.mes-hall]}
  \BibitemShut {NoStop}%
\bibitem [{\citenamefont {Grushin}(2012)}]{Grushin:2012mt}%
  \BibitemOpen
  \bibfield  {author} {\bibinfo {author} {\bibfnamefont {A.~G.}\ \bibnamefont
  {Grushin}},\ }\href {https://doi.org/10.1103/PhysRevD.86.045001} {\bibfield
  {journal} {\bibinfo  {journal} {Phys. Rev. D}\ }\textbf {\bibinfo {volume}
  {86}},\ \bibinfo {pages} {045001} (\bibinfo {year} {2012})},\ \Eprint
  {https://arxiv.org/abs/1205.3722} {arXiv:1205.3722 [hep-th]} \BibitemShut
  {NoStop}%
\bibitem [{\citenamefont {Kharzeev}(2014)}]{Kharzeev:2013ffa}%
  \BibitemOpen
  \bibfield  {author} {\bibinfo {author} {\bibfnamefont {D.~E.}\ \bibnamefont
  {Kharzeev}},\ }\href {https://doi.org/10.1016/j.ppnp.2014.01.002} {\bibfield
  {journal} {\bibinfo  {journal} {Prog. Part. Nucl. Phys.}\ }\textbf {\bibinfo
  {volume} {75}},\ \bibinfo {pages} {133} (\bibinfo {year} {2014})},\ \Eprint
  {https://arxiv.org/abs/1312.3348} {arXiv:1312.3348 [hep-ph]} \BibitemShut
  {NoStop}%
\bibitem [{\citenamefont {Wilczek}(1987)}]{Wilczek:1987mv}%
  \BibitemOpen
  \bibfield  {author} {\bibinfo {author} {\bibfnamefont {F.}~\bibnamefont
  {Wilczek}},\ }\href {https://doi.org/10.1103/PhysRevLett.58.1799} {\bibfield
  {journal} {\bibinfo  {journal} {Phys. Rev. Lett.}\ }\textbf {\bibinfo
  {volume} {58}},\ \bibinfo {pages} {1799} (\bibinfo {year}
  {1987})}\BibitemShut {NoStop}%
\bibitem [{\citenamefont {Carroll}\ \emph {et~al.}(1990)\citenamefont
  {Carroll}, \citenamefont {Field},\ and\ \citenamefont
  {Jackiw}}]{Carroll:1989vb}%
  \BibitemOpen
  \bibfield  {author} {\bibinfo {author} {\bibfnamefont {S.~M.}\ \bibnamefont
  {Carroll}}, \bibinfo {author} {\bibfnamefont {G.~B.}\ \bibnamefont {Field}},\
  and\ \bibinfo {author} {\bibfnamefont {R.}~\bibnamefont {Jackiw}},\ }\href
  {https://doi.org/10.1103/PhysRevD.41.1231} {\bibfield  {journal} {\bibinfo
  {journal} {Phys. Rev. D}\ }\textbf {\bibinfo {volume} {41}},\ \bibinfo
  {pages} {1231} (\bibinfo {year} {1990})}\BibitemShut {NoStop}%
\bibitem [{\citenamefont {Sikivie}(1984)}]{Sikivie:1984yz}%
  \BibitemOpen
  \bibfield  {author} {\bibinfo {author} {\bibfnamefont {P.}~\bibnamefont
  {Sikivie}},\ }\href {https://doi.org/10.1016/0370-2693(84)91731-3} {\bibfield
   {journal} {\bibinfo  {journal} {Phys. Lett. B}\ }\textbf {\bibinfo {volume}
  {137}},\ \bibinfo {pages} {353} (\bibinfo {year} {1984})}\BibitemShut
  {NoStop}%
\bibitem [{\citenamefont {Arnold}\ \emph {et~al.}(1997)\citenamefont {Arnold},
  \citenamefont {Son},\ and\ \citenamefont {Yaffe}}]{Arnold:1996dy}%
  \BibitemOpen
  \bibfield  {author} {\bibinfo {author} {\bibfnamefont {P.~B.}\ \bibnamefont
  {Arnold}}, \bibinfo {author} {\bibfnamefont {D.}~\bibnamefont {Son}},\ and\
  \bibinfo {author} {\bibfnamefont {L.~G.}\ \bibnamefont {Yaffe}},\ }\href
  {https://doi.org/10.1103/PhysRevD.55.6264} {\bibfield  {journal} {\bibinfo
  {journal} {Phys. Rev. D}\ }\textbf {\bibinfo {volume} {55}},\ \bibinfo
  {pages} {6264} (\bibinfo {year} {1997})},\ \Eprint
  {https://arxiv.org/abs/hep-ph/9609481} {arXiv:hep-ph/9609481} \BibitemShut
  {NoStop}%
\bibitem [{\citenamefont {Arnold}\ \emph {et~al.}(1999)\citenamefont {Arnold},
  \citenamefont {Son},\ and\ \citenamefont {Yaffe}}]{Arnold:1998cy}%
  \BibitemOpen
  \bibfield  {author} {\bibinfo {author} {\bibfnamefont {P.~B.}\ \bibnamefont
  {Arnold}}, \bibinfo {author} {\bibfnamefont {D.~T.}\ \bibnamefont {Son}},\
  and\ \bibinfo {author} {\bibfnamefont {L.~G.}\ \bibnamefont {Yaffe}},\ }\href
  {https://doi.org/10.1103/PhysRevD.59.105020} {\bibfield  {journal} {\bibinfo
  {journal} {Phys. Rev. D}\ }\textbf {\bibinfo {volume} {59}},\ \bibinfo
  {pages} {105020} (\bibinfo {year} {1999})},\ \Eprint
  {https://arxiv.org/abs/hep-ph/9810216} {arXiv:hep-ph/9810216} \BibitemShut
  {NoStop}%
\bibitem [{\citenamefont {Bodeker}(1998)}]{Bodeker:1998hm}%
  \BibitemOpen
  \bibfield  {author} {\bibinfo {author} {\bibfnamefont {D.}~\bibnamefont
  {Bodeker}},\ }\href {https://doi.org/10.1016/S0370-2693(98)00279-2}
  {\bibfield  {journal} {\bibinfo  {journal} {Phys. Lett. B}\ }\textbf
  {\bibinfo {volume} {426}},\ \bibinfo {pages} {351} (\bibinfo {year}
  {1998})},\ \Eprint {https://arxiv.org/abs/hep-ph/9801430}
  {arXiv:hep-ph/9801430} \BibitemShut {NoStop}%
\bibitem [{\citenamefont {Akamatsu}\ and\ \citenamefont
  {Yamamoto}(2013)}]{Akamatsu:2013pjd}%
  \BibitemOpen
  \bibfield  {author} {\bibinfo {author} {\bibfnamefont {Y.}~\bibnamefont
  {Akamatsu}}\ and\ \bibinfo {author} {\bibfnamefont {N.}~\bibnamefont
  {Yamamoto}},\ }\href {https://doi.org/10.1103/PhysRevLett.111.052002}
  {\bibfield  {journal} {\bibinfo  {journal} {Phys. Rev. Lett.}\ }\textbf
  {\bibinfo {volume} {111}},\ \bibinfo {pages} {052002} (\bibinfo {year}
  {2013})},\ \Eprint {https://arxiv.org/abs/1302.2125} {arXiv:1302.2125
  [nucl-th]} \BibitemShut {NoStop}%
\bibitem [{\citenamefont {Manuel}\ and\ \citenamefont
  {Torres-Rincon}(2014)}]{Manuel:2013zaa}%
  \BibitemOpen
  \bibfield  {author} {\bibinfo {author} {\bibfnamefont {C.}~\bibnamefont
  {Manuel}}\ and\ \bibinfo {author} {\bibfnamefont {J.~M.}\ \bibnamefont
  {Torres-Rincon}},\ }\href {https://doi.org/10.1103/PhysRevD.89.096002}
  {\bibfield  {journal} {\bibinfo  {journal} {Phys. Rev. D}\ }\textbf {\bibinfo
  {volume} {89}},\ \bibinfo {pages} {096002} (\bibinfo {year} {2014})},\
  \Eprint {https://arxiv.org/abs/1312.1158} {arXiv:1312.1158 [hep-ph]}
  \BibitemShut {NoStop}%
\bibitem [{\citenamefont {Peskin}\ and\ \citenamefont
  {Schroeder}(1995)}]{Peskin:1995ev}%
  \BibitemOpen
  \bibfield  {author} {\bibinfo {author} {\bibfnamefont {M.~E.}\ \bibnamefont
  {Peskin}}\ and\ \bibinfo {author} {\bibfnamefont {D.~V.}\ \bibnamefont
  {Schroeder}},\ }\href@noop {} {\emph {\bibinfo {title} {{An Introduction to
  quantum field theory}}}}\ (\bibinfo  {publisher} {Addison-Wesley},\ \bibinfo
  {address} {Reading, USA},\ \bibinfo {year} {1995})\BibitemShut {NoStop}%
\bibitem [{\citenamefont {Rischke}(2004)}]{RISCHKE2004197}%
  \BibitemOpen
  \bibfield  {author} {\bibinfo {author} {\bibfnamefont {D.~H.}\ \bibnamefont
  {Rischke}},\ }\href
  {https://doi.org/https://doi.org/10.1016/j.ppnp.2003.09.002} {\bibfield
  {journal} {\bibinfo  {journal} {Progress in Particle and Nuclear Physics}\
  }\textbf {\bibinfo {volume} {52}},\ \bibinfo {pages} {197} (\bibinfo {year}
  {2004})}\BibitemShut {NoStop}%
\bibitem [{\citenamefont {Cao}\ and\ \citenamefont {Wang}(2021)}]{Cao_2021}%
  \BibitemOpen
  \bibfield  {author} {\bibinfo {author} {\bibfnamefont {S.}~\bibnamefont
  {Cao}}\ and\ \bibinfo {author} {\bibfnamefont {X.-N.}\ \bibnamefont {Wang}},\
  }\href {https://doi.org/10.1088/1361-6633/abc22b} {\bibfield  {journal}
  {\bibinfo  {journal} {Reports on Progress in Physics}\ }\textbf {\bibinfo
  {volume} {84}},\ \bibinfo {pages} {024301} (\bibinfo {year}
  {2021})}\BibitemShut {NoStop}%
\end{thebibliography}%

\end{document}